\documentclass[12pt]{article}
\usepackage{epsf}
\begin{document}
\title{Conductivity magnetooscillations in 2D electron-impurity system under microwave irradiation: role of magnetoplasmons}
\author{{\em E.~E.~Takhtamirov} and V.~A.~Volkov\\
Institute of Radioengineering and Electronics of RAS\\
125009 Moscow, Russia}
\date{}
\maketitle

\begin{abstract}
{
It is developed a many-electron approach to explain the recently observed conductivity magnetooscillations in very high mobility 2D electron systems under microwave irradiation. For the first time a theory takes into account the microwave-induced renormalization of the screened impurity potential. As a result this potential has singular, dynamic and non-linear in electric field nature. That changes the picture of scattering of electrons at impurities in a ``clean'' 2D system essentially: for appearence of the rectified dissipative current responsible are excitations of 2D magnetoplasmons rather than one-electron transitions between Landau levels. In a ``dirty'' 2D system the role of electron-electron interaction diminishes, so the collective excitations cease to exist, and our results turn into the well-known ones, which were obtained in the one-electron approach.
}
\end{abstract}

For a high quality 2D electron system in structures Ga\-As\-/Al\-Ga\-As subjected to microwave (MW) field with frequency $\Omega$ it was found  that the magnetoresistance experienced oscillations governed with the ratio $\Omega / \omega_c$ \cite{Zudov1}, where $\omega_c = e B/(m^*c)$ is the cyclotron frequency. The states with zero resistance were observed with an increase of MW field intensity \cite{Mani}. These observations have been confirmed by other researchers, see the review \cite{Du}. That brought about an ava\-lan\-che of theoretical works.

There exist two mainstream theoretical scenarios of the effect, both being one-electron. The first one is based on the mechanism of electron displacement against strong external DC field as a result of MW absorption and impurity scattering, and this was shown to be capable of leading to the absolute negative DC conductivity \cite{Ryzhiietall}. Being then unstable, the system breaks into domains, and one just registers zero resistance \cite{unstable}. The second scenario is based on the wave-induced inversion of electron population on higher Landau levels (LL) \cite{Dorozhkin}. Indisputable explanation of the main experimental data has not been achieved yet.

In this work we consider the effect of electron-electron (e-e) interaction on the impurity scattering confining ourselves to the first scenario. Here we analyse the case of an unbounded high-quality 2D electron system with very weak impurity scattering. The main obvious consequence of e-e interaction is the screening of the impurity potential with 2D electrons. At first glance it seems that e-e interaction is not able to induce a qualitative change in the results of Ref.~\cite{Ryzhiietall}. But it is shown to be a delusion. With that, for the very appearence of the dissipative direct current responsible are not the usually considered single-particle transitions of electrons between LL, but rather 2D magnetoplasmons. Below in the framework of the random phase approximation (RPA) it is developed a systematic theory of the non-liner dissipative conductivity applicable to the experimental conditions \cite{Zudov1,Mani}.

Dealing with our system, let us change the reference frame to the one connected with the external homogenious electric field. In such a reference frame electrons do not experience the external electric field if no impurity is available in the system. Being presented and so transformed, the bare potential of the impurity system becomes time-dependent:
\begin{equation}
V_{\rm imp} \left( {\bf r} \right) \rightarrow
V_{\rm imp} \left( {\bf r} - {\bf r}_0 \left( t \right) \right),
\label{potentialtransform}
\end{equation}
where ${\bf r}_0 \left( t \right)$ is the radius-vector describing movement of the center of the classical oscillator in external electric field. To screen the transformed potential, which is the right-hand side of the transform (\ref{potentialtransform}), one should use the dynamic dielectric function. The space-time Fourier transform of the screened potential is
\begin{equation}
V^{\rm (scr)}_{\rm imp} \left( {\bf q},\omega \right) =
\frac {V_{\rm imp} \left( {\bf q}, \omega \right)} {\varepsilon \left( q, \omega \right) },
\label{screenedpotential}
\end{equation}
where $V_{\rm imp} \left( {\bf q}, \omega \right)$ is the Fourier image of the right-hand side of the transform (\ref{potentialtransform}), and $\varepsilon \left( q, \omega \right)$  is the dielectric function. Being obtained in RPA, it has the form:
\begin{eqnarray}
\varepsilon \left( q, \omega \right) = 1+ \frac {V_{\rm ee}(q)}{\pi \hbar \lambda^2}\sum_{M,M'} \frac {\left(f_{M} - f_{M'}\right)I_{M,M'}\left(q\right)} {\omega_c \left(M'-M\right) +\omega +i0}, \nonumber
\end{eqnarray}
where $V_{\rm ee}(q)$ is the Fourier transform of potential of e-e interaction, for 2D electron gas in a medium with the constant lattice dielectric permeability $\kappa$ we have $V_{\rm ee}(q) = 2\pi e^2/(\kappa q)$, $\lambda = \sqrt{\hbar c/\left(e B\right)}$ is the magnetic length, $f_{M}$ is the Fermi distribution function, $M$ and $M'$ are LL indices. And
$I_{M,M'}\left(q\right)$ is the square of the absolute value of the overlap integral of the Landau functions with the oscillator centers shifted by $q\lambda^2$. At the magnetoplasmon frequency $\omega = \omega_{\rm MP}(q)$ the denominator in Eq.~(\ref{screenedpotential}) turns to zero, so the screening in the strong field does not ordinarily soften the impurity potental, but rather sharply strengthen it. Let the external homogeneous electric field ${\bf F}$ be a sum of AC field of the wave with the amplitude $W$ and DC dragging field $F_{\rm DC}$, both having only one ($x$) component for simplicity:
\begin{eqnarray}
F_x = F_{\rm DC} + W\sin \Omega t. \nonumber
\end{eqnarray}
The current density in the system is ${\bf j} = -en_{\rm s} {\rm Tr}(\rho {\bf v})$, where $n_{\rm s}$ is 2D electron concentration, ${\bf v}$ is the velocity operator, $\rho$ is the density matrix that meets the quantum kinetic equation. Following Ref.~\cite{Adams} we solve the kinetic equation at low order in scattering of electrons at the screened impurities. After taking an average of all chaotic impurity configurations, asuming only one type of impurity with 2D concentration $n_{\rm imp}$ and the bare single impurity potential $V^{(0)}_{\rm imp} \left( q \right)$, we have for the time average of the dissipative current density:
\begin{eqnarray}
\langle j_x \rangle &=&  - \frac {e n_{\rm imp}} {\left( 2\pi \right)^2 m^* \omega_c } \int {\rm d}^2 q  \frac { \mid V^{(0)}_{\rm imp}( q)\mid^2} {V_{\rm ee}(q)} q_y \nonumber\\
&&\times \sum_{n=-\infty}^{+\infty} J_n^2(Q) \, {\rm Im} \varepsilon^{-1} \left(q, q_y v_{\rm H} + n \Omega \right), \label{jxeps}
\end{eqnarray}
where $J_n$ is the Bessel function, $Q = \left(Q_x^2+Q_y^2\right)^{1/2}$,
\begin{eqnarray}
Q_x = \frac {q_x eW} {m^* (\omega_c^2 - \Omega^2)},\quad Q_y  = \frac {q_ye W\omega_c} {m^*\Omega (\omega_c^2 - \Omega^2)}, \nonumber
\end{eqnarray}
$v_{\rm H} = c F_{\rm DC}/B$ is the Hall velocity.
If we neglect the colli\-si\-on-induced LL broading,
\begin{equation}
{\rm Im} \frac 1 {\varepsilon \left(q, q_y v_{\rm H} + n \Omega \right)} =
-\pi \sum_p \frac {\delta \left(q_y v_{\rm H} + n \Omega -\omega _p \right)}
{\varepsilon ^{\prime}_\omega \left(q,\omega_p \right)}, \label{imeps}
\end{equation}
where $\omega_p = \omega_p \left( q \right)$, index $p = \pm 1,\, \pm 2, \ldots$ enumerates all solutions to the dispersion equation $\varepsilon \left(q,\omega_p \right) = 0$, so that $\omega_p \to p \omega_c$ as $q \to \infty$, and $\varepsilon^{\prime}_\omega \left(q,\omega_p \right) = {\rm d}\varepsilon \left(q,\omega \right)/{\rm d} \omega \bigr| _{\omega = \omega _p}$.

When no MW field is given, $W=0$, only the term with $n=0$ survives in the sum of (\ref{jxeps}), $J_0(0) = 1$. In such a form our result, which generalizes the one of Ref.~\cite{Tavger} obtained with leaving e-e interaction out, is applicable to explanation of the experiment \cite{Yang}.

With the help of polar coordinates in Eq.~(\ref{jxeps}): $q_x = q \cos {\phi}$, $q_y = q \sin {\phi}$, using Eq.~(\ref{imeps}) and integrating by ${\rm d} \phi$ we obtain the expression that allows graphical analysis:
\begin{eqnarray}
\langle j_x \rangle  &=&  \frac {e n_{\rm imp}} {2\pi m^* \omega_c} \int_0^{+\infty}{\rm d} q \frac { \mid V^{(0)}_{\rm imp}( q)\mid^2} {V_{\rm ee}(q)} \sum_p \frac {q^2} {\varepsilon ^{\prime}_\omega \left(q,\omega_p \right)} \nonumber\\
&&\times \sum_{n=-\infty}^{+\infty}J_n^2(\bar Q_p) \frac {\omega_p - n\Omega} {q v_{\rm H}}
\frac 1 {\sqrt{ q^2 v^2_{\rm H} - (\omega_p - n\Omega)^2 }} \label{jxpol} \\
&&\times \left(
\Theta \left(\frac {\omega_p - n\Omega} {q v_{\rm H}} +1 \right) -
\Theta \left(\frac {\omega_p - n\Omega} {q v_{\rm H}} - 1 \right)
\right).\nonumber
\end{eqnarray}
Here $\Theta $ is the Heaviside step-function, and
\begin{eqnarray}
\bar Q_p = \frac {qeW} {m^* (\omega_c^2 - \Omega^2)} \sqrt{
1+ \left( \frac {\omega_c^2} {\Omega^2} - 1 \right)
\left( \frac {\omega_p - n\Omega} {q v_{\rm H}} \right)^2 
},\nonumber
\end{eqnarray}
In Fig.~1 shown are the spectrum of the principal magnetoplasmon (at $p = 1$ and $\omega_p(q) > 0$), two lines ($\omega = n \Omega \pm q v_{\rm H}$) forming a region that confines all values of $\omega_p$ and so $q$ contributing to the integral (see the last line of Eq.~(\ref{jxpol})), and the bisector ($\omega = n \Omega$) parting that region onto two ones contributing purely positive or negative.
\begin{figure}[ht]
\leavevmode \centering{\epsfysize=71mm \epsfbox{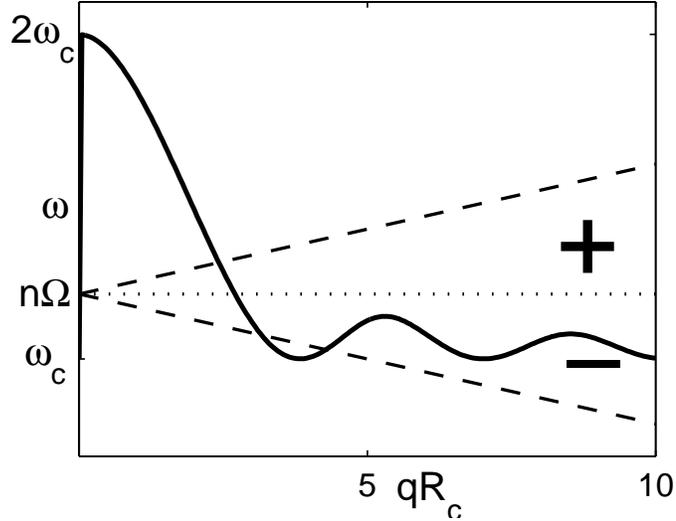}}
\caption[]{Full thick line: spectrum of the principal magnetoplasmon; dashed lines: boundaries $\omega = n \Omega \pm q v_{\rm H}$ of the contribution region; dotted line, $\omega = n \Omega$, divides the region onto the ones of positive `` $+$'' and negative ``$-$'' contributions to dissipative current; $R_c$ is the Larmour radius.}
\end{figure}
Let $F_{\rm DC} > 0$ be weak enough (say $v_{\rm H} < \omega_c/2k_{\rm F}$, where $\hbar k_{\rm F}$ is the Fermi momentum), so that the term with $n=0$ in the sum of Eq.~(\ref{jxeps}) does not play a role with its always positive contribution. And let us consider one-photon processes only: $n=\pm 1$ in the sum of Eq.~(\ref{jxeps}). Then $\langle j_x \rangle > 0$ if $\Omega < \omega_c$, that meets the positive magnetoresistance. Other case, if $\omega < \Omega < 2\omega_c$, a bunch of magnetoplasmon modes may fall into the region of negative contribution. That gives the absolute negative conductivity. Similar picture holds for higher values $\Omega$, and the higher magnetoplasmon modes $|p|>1$ take part in the play. In contrast to theories omitting the screening \cite{Ryzhiietall}, the regions of positive and negative conductivities are finite even in an ideal case of no LL broading (\ref{imeps}), and anyhow small $F_{\rm DC}$ be. 

In a ``dirty'' 2D system the role of electron-electron interaction diminishes. It is somewhat equivalent to $V_{\rm ee} \to 0$. Then in the vicinity of $\omega = \omega_p(q) \approx p\omega_c$ we may use the approximate expression for $\varepsilon (q, \omega )$:
\begin{eqnarray}
\varepsilon_p \left( q,\omega \right) = 1+ \frac {2m^*V_{\rm ee}(q)}{\pi \hbar^2} \frac {p^2\omega_c^2}
{p^2\omega_c^2 - \omega^2 - i0{\rm sign}\omega} \bar I_p\left(q\right),
\nonumber
\end{eqnarray}
where
\begin{eqnarray}
\bar I_p\left(q\right) = p^{-1}\sum_{M=0}^{\infty} {\left(f_{M} - f_{M +p}\right)I_{M,M+p}\left(q\right)}. \nonumber
\end{eqnarray}
Then $V_{\rm ee}$ falls out, and with condition (\ref{imeps}), Eq.~(\ref{jxeps}) transforms to
\begin{eqnarray}
\langle j_x \rangle &=&  \frac {e n_{\rm imp}} {\left( 2\pi \hbar \right)^2} \int {\rm d}^2 q  \mid V^{(0)}_{\rm imp}( q)\mid^2 q_y \sum_{n=-\infty}^{+\infty} J_n^2(Q) \nonumber\\
&&\times \sum_p p \bar I_p\left(q\right)\,\delta \left(q_y v_{\rm H} + n \Omega -p\omega _c \right), \nonumber
\end{eqnarray}
which is the result of Ref.~\cite{Ryzhiietall} for ideal infinitely narrow LL, and the result of Ref.~\cite{Tavger} for the case of DC field only.

The work was supported by RFBR and RAS programme.

\end{document}